\documentstyle[aps,preprint,floats,epsf]{revtex}
\tightenlines

\def\beq{\begin{equation}}
\def\eeq{\end{equation}}
\def\bea{\begin{eqnarray}}
\def\eea{\end{eqnarray}}

\begin{document}

\thispagestyle{empty}

\font\fortssbx=cmssbx10 scaled \magstep2
\hbox to \hsize{
\hbox{\fortssbx University of Wisconsin - Madison}
      \hfill$\vcenter{
\hbox{\bf FERMILAB-PUB-98-385-T}
\hbox{\bf MADPH-98-1080}
\hbox{\bf AMES-HET-98-13}
       \hbox{November 1998}}$ }

\vspace{.5in}

\begin{center}
{\bf GLOBAL THREE-NEUTRINO VACUUM OSCILLATION FITS\\
TO THE SOLAR AND ATMOSPHERIC ANOMALIES}\\
\vskip 0.7cm
{V. Barger$^{1,2}$ and K. Whisnant$^3$}
\\[.1cm]
$^1${\it Fermi National Accelerator Laboratory, Batavia, IL 60510 USA}\\
$^2${\it Department of Physics, University of Wisconsin, Madison, WI
53706 USA}\\
$^3${\it Department of Physics and Astronomy, Iowa State University,
Ames, IA 50011 USA}\\
\end{center}

\smallskip

\begin{abstract}

We determine the three-neutrino mixing and mass parameters that are
allowed by the solar and atmospheric neutrino data when vacuum
oscillations are responsible for both phenomena. The global fit does not
appreciably change the allowed regions for the parameters obtained from
effective two-neutrino fits. We discuss how measurements of the solar
electron energy spectrum below 6.5~GeV in Super-Kamiokande and seasonal
variations in the Super-Kamiokande, $^{71}$Ga, and BOREXINO experiments
can distinguish the different solar vacuum solutions.

\end{abstract}

\thispagestyle{empty}
\newpage

\section{Introduction}

Recent data from the Super-Kamiokande 
experiment~\cite{SuperKsolar,SuperKatmos} have strengthened the
interpretation of the solar~\cite{solar1,solar2,solar3,SSM} and
atmospheric~\cite{oldatmos,atmos} neutrino anomalies in terms of
neutrino oscillations. Oscillations have also been invoked to describe
the appearance of electron neutrinos and antineutrinos in the LSND
experiment~\cite{LSND}.  Because confirmation of the LSND results awaits
future experiments and recent measurements in the KARMEN detector
exclude part of the LSND allowed region~\cite{KARMEN}, a conservative
approach is to assume that oscillations need only account for the solar
and atmospheric data. Then the two mass-squared difference scales in a
three-neutrino model are sufficient to describe the data. A number of
three-neutrino models have been
investigated~\cite{newatmos,rpv,bimax,maxosc,hall,newmodels}. One
attractive possibility is that both the atmospheric $\nu_\mu$ and solar
$\nu_e$ oscillate maximally~\cite{bimax} or
near-maximally~\cite{maxosc} at the $\delta m^2_{atm}$ and $\delta
m^2_{sun}$ scales, respectively.

Most detailed oscillation fits have been done separately for the solar
and atmospheric data in effective two-neutrino
approximations~\cite{global}. In this paper we make full
three-neutrino fits to the solar and atmospheric oscillation data to
determine the allowed values for the general three-neutrino mixing
matrix under the assumption that one mass-squared difference, $\delta
m^2_{atm}$, explains the atmospheric neutrino oscillations and that the
other independent mass-squared difference, $\delta m^2_{sun} \ll \delta
m^2_{atm}$, explains the solar neutrino oscillations via the vacuum
long-wavelength scenario~\cite{vlw,vlw2}. We choose a particular
parametrization for the three-neutrino mixing in which the expressions
for the atmospheric and solar neutrino oscillations share only a single
parameter, and discuss the degree to which the two phenomena decouple.
We find that the extension from two to three neutrino species does not
improve the separate fits to either the atmospheric or solar
data. Although pure $\nu_\mu \rightarrow \nu_\tau$ oscillations of
atmospheric neutrinos are favored, there exist three-neutrino solutions
with non-negligible $\nu_\mu \leftrightarrow \nu_e$ oscillations, even
with the constraints from the CHOOZ reactor
experiment~\cite{CHOOZ}. Hence both $\nu_\mu \leftrightarrow \nu_e$ and
$\nu_e \rightarrow \nu_\tau$ oscillations may be observable in future
long-baseline experiments. We also find that future measurements of
seasonal variations in the solar neutrino signal can provide an
unmistakable sign of vacuum oscillations, and can further constrain the
allowed parameter regions.

In Sec.~II we review the formalism for oscillations of three neutrinos
relevant to the atmospheric and solar phenomena.  In Sec.~III we evaluate
the current allowed two-neutrino parameter regions, and briefly review
the evidence indicating that two distinct $\delta m^2$ are needed to
describe the solar and atmospheric data. In Sec.~IV we obtain
three-neutrino solutions from a combined fit to the atmospheric and
solar data. We conclude with a discussion of future tests of vacuum
oscillations in Sec.~V.

\section{Oscillation analysis}

\subsection{General probability expressions}

The survival probability for a given neutrino flavor $\nu_\alpha$ in a
vacuum is~\cite{VBreal}
\beq
P(\nu_\alpha\to \nu_\alpha) = 1
- 4 \sum_{k<j} |U_{\alpha j}|^2 |U_{\alpha k}|^2 \sin^2 \Delta_{jk} \,,
\label{oscprob}
\eeq
where $U$ is the neutrino mixing matrix (in the basis where the
charged-lepton mass matrix is diagonal), $\Delta_{jk} \equiv \delta
m_{jk}^2 \,L/4E = 1.27 (\delta m^2_{jk}/{\rm eV}^2) (L/{\rm km})/(E/{\rm
GeV})$, $\delta m^2_{jk}\equiv m^2_j-m^2_k$, and the sum is over all $j$
and $k$, subject to $k<j$. The matrix elements $U_{\alpha j}$ are the
mixings between the flavor ($\alpha=e,\mu,\tau$) and the mass
($j=1,2,3$) eigenstates, and we assume without loss of generality that
$m_1 < m_2 < m_3$. The solar oscillations are driven by
$|\Delta_{21}| \equiv \Delta_{sun}$ and the atmospheric oscillations are
driven by $|\Delta_{31}| \simeq |\Delta_{32}| \equiv \Delta_{atm} \gg
\Delta_{sun}$.

The off-diagonal vacuum oscillation probabilities in a three-neutrino
model are~\cite{bww98}
\bea
P(\nu_e\rightarrow\nu_\mu) &=&
4\,|U_{e3} U_{\mu 3}^*|^2 \sin^2\Delta_{atm}
-4 Re\{ U_{e1} U_{e2}^* U_{\mu 1}^* U_{\mu 2}\} \sin^2\Delta_{sun}
-2\,J\sin 2\Delta_{sun}\,,
\label{pemu} \\
P(\nu_e\rightarrow\nu_\tau) &=&
4\,|U_{e3} U_{\tau 3}^*|^2 \sin^2\Delta_{atm}
-4 Re\{U_{e1} U_{e2}^* U_{\tau 1}^* U_{\tau 2}\} \sin^2\Delta_{sun}
+2\,J\sin 2\Delta_{sun}\,,
\label{petau} \\
P(\nu_\mu\rightarrow\nu_\tau) &=&
4\,|U_{\mu 3} U_{\tau 3}^*|^2 \sin^2\Delta_{atm}
-4 Re\{ U_{\mu 1} U_{\mu 2}^* U_{\tau 1}^* U_{\tau 2}\} \sin^2\Delta_{sun}
-2\,J\sin 2\Delta_{sun}\,,
\label{pmutau}
\eea
where the $CP$-violating ``Jarlskog invariant''~\cite{jarlskog}
is $J = \sum_{k,\gamma} \epsilon_{ijk} \epsilon_{\alpha\beta\gamma}
Im\{ U_{\alpha i} U_{\alpha j}^* U_{\beta i}^* U_{\beta j}\}$ for any
$\alpha$, $\beta$, $i$, and $j$ (e.g., $J = Im\{ U_{e2} U_{e3}^*
U_{\mu 2}^* U_{\mu 3}\}$ for $\alpha=e$, $\beta=\mu$, $i=2$, and $j=3$).
The $CP$-odd term changes sign under reversal of the oscillating
flavors, or if neutrinos are replaced by anti-neutrinos.
We note that the $CP$-violating probability at the atmospheric scale is
suppressed to order $\delta m^2_{sun}/\delta m^2_{atm}$, with the leading
term cancelling in the sum over the two light-mass states; thus,
$P(\nu_\alpha \rightarrow \nu_\beta) = P(\bar\nu_\alpha \rightarrow
\bar\nu_\beta)$ (and therefore $P(\nu_\alpha \rightarrow \nu_\beta) =
P(\nu_\beta \rightarrow \nu_\alpha)$ from $CPT$ invariance) at the
atmospheric scale.

Without loss of generality we work in the basis where the
charged-lepton mass matrix is diagonal.  The matrix $U$ that relates
the flavor eigenstates to the mass eigenstates may be parametrized in
terms of three Euler angles and one (three) phases for Dirac (Majorana)
neutrinos. This can be understood as follows. In the Dirac case, $U$ is
analogous to the CKM mixing matrix in the quark sector, in which there
are three Euler angles and one phase.  For Majorana neutrinos, the
$3\times3$ mass matrix in the flavor basis must be symmetric but may be
complex~\cite{mopal}. There are 12 independent parameters, which can be
taken as 3 real mass eigenvalues, 3 real Euler angles, and 6
phases. Then three of the phases can be absorbed into the definitions of
the fields. However, only one of the three remaining phases has physical
consequences in neutrino oscillations. Therefore for either Dirac or
Majorana neutrinos we can choose the following parametrization for
$U$~\cite{PDG}
\beq
\left( \begin{array}{c} \nu_e \\ \nu_\mu \\ \nu_\tau \end{array} \right)
= U \left( \begin{array}{c} \nu_1 \\ \nu_2 \\ \nu_3 \end{array} \right)
= \left( \begin{array}{ccc}
  c_1 c_3                           & c_1 s_3
& s_1 e^{-i\delta} \\
- c_2 s_3 - s_1 s_2 c_3 e^{i\delta} &   c_2 c_3 - s_1 s_2 s_3 e^{i\delta}
& c_1 s_2 \\
  s_2 s_3 - s_1 c_2 c_3 e^{i\delta} & - s_2 c_3 - s_1 c_2 s_3 e^{i\delta}
& c_1 c_2 \\
\end{array} \right)
\left( \begin{array}{c}
\nu_1 \\ \nu_2 \\ \nu_3
\end{array} \right) \,,
\label{U}
\eeq
where $c_j \equiv \cos\theta_j$, $s_j \equiv \sin\theta_j$ and
$\delta$ is the $CP$-violating phase.

\subsection{Atmospheric and long-baseline experiments}

The experimental indications are that $\delta m^2_{atm} \sim
10^{-3}$~eV$^2$ and that $\delta m^2_{sun} \sim 10^{-10}$~eV$^2$ for a
vacuum oscillation explanation of the solar neutrino data. Then for the
oscillations of neutrinos in atmospheric and long-baseline experiments
with $L/E \agt 10^2$~km/GeV, the $\Delta_{sun}$ terms are negligible
and the relevant oscillation probabilities are
\bea
P(\nu_\mu\rightarrow\nu_\mu) &=&
1 - (c_1^4 \sin^22\theta_2 + s_2^2 \sin^22\theta_1) \sin^2\Delta_{atm} \,,
\label{pmumu2} \\
P(\nu_e\rightarrow\nu_e) &=& 1 -
\sin^22\theta_1 \sin^2\Delta_{atm} \,.
\label{pee2} \\
P(\nu_e\leftrightarrow\nu_\mu) &=&
s_2^2 \sin^22\theta_1 \sin^2\Delta_{atm} \,.
\label{pemu2} \\
P(\nu_e\leftrightarrow\nu_\tau) &=&
c_2^2 \sin^2 2\theta_1 \sin^2\Delta_{atm} \,,
\label{petau2} \\
P(\nu_\mu\leftrightarrow\nu_\tau) &=&
c_1^4 \sin^2 2\theta_2 \sin^2\Delta_{atm} \,.
\label{pmutau2}
\eea
When $\theta_1=0$ (i.e., $U_{e3}=0$), these reduce to pure $\nu_\mu
\rightarrow \nu_\tau$ oscillations with amplitude $\sin^22\theta_2$, and
$\nu_e$ does not oscillate in atmospheric and long-baseline experiments.

We define the oscillation amplitudes $A^{\mu\not\mu}_{atm}$, $A^{e \not
e}_{atm}$, $A^{\mu e}_{atm}$, $A^{e\tau}_{atm}$, and $A^{\mu\tau}_{atm}$
as the coefficients of the $\sin^2 \Delta_{atm}$ terms in
Eqs.~(\ref{pmumu2})--(\ref{pmutau2}), respectively. The neutrino
parameters can then be determined from the atmospheric neutrino data by
the relations
\beq
N_\mu/N^o_\mu = \alpha \left[(1 - \left< S \right> \, A^{\mu\not\mu}_{atm})
+ r \, \left< S \right> \, A^{\mu e}_{atm} \right] \,,
\label{Rmu}
\eeq
and
\beq
N_e/N^o_e = \alpha \left[ (1 - \left< S \right> \, A^{e\not e}_{atm})
+ r^{-1} \, \left< S \right> \, A^{\mu e}_{atm} \right] \,,
\label{Re}
\eeq
where $N_e^o$ and $N_\mu^o$ are the expected numbers of atmospheric $e$
and $\mu$ events, respectively, $r\equiv N^o_e/N^o_\mu$, $\left< S
\right>$ is $\sin^2\Delta_{atm}$ appropriately averaged,
and $\alpha$ is an overall neutrino flux normalization, which we allow
to vary following the SuperK analysis~\cite{SuperKatmos}.

SuperK presented $N_{\mu}/N_{\mu}^o$ and $N_e/N_e^o$ for eight different
$L/E$ bins~\cite{SuperKatmos} from a 535 day exposure. The data were
obtained by inferring an $L/E$ value for each event from the zenith
angle $\theta_\ell$ and energy of the observed charged lepton $E_\ell$
and comparing it to expectations from a monte carlo simulation based on
the atmospheric neutrino spectrum~\cite{flux} folded with the
differential cross section.  Due to the fact that the charged lepton
energy and direction in general differ from the corresponding values for
the incident neutrino (or antineutrino), the $L/E$ distribution involves
substantial smearing. We estimate this smearing by a monte carlo
integration over the neutrino angle and energy
spectrum~\cite{stanevflux} weighted by the deep-inelastic differential cross section\cite{collphys}.
We generate events with $E_\nu$ and $\theta_\nu$, and determine
the corresponding $E_\ell$ and $\theta_\ell$ for the charged lepton. We
bin the events in $L/E_\nu$, using $\theta_\ell$ to determine $L$ and an
estimated neutrino energy inferred from the average ratio of lepton
momentum to neutrino energy, $E_\nu^{\rm est} = E_\ell \left<
E_\nu/E_\ell \right>$, analogous to the SuperK
analysis~\cite{SuperKatmos}. We calculate a value for $\left<
\sin^2\Delta_{atm} \right>$ for each $L/E$ bin for a given value of
$\delta m^2_{atm}$. We can then fit Eqs.~(\ref{Rmu}) and (\ref{Re}) to
the data and determine the neutrino parameters in
Eqs.~(\ref{pmumu2})--(\ref{pemu2}) and the normalization
$\alpha$. Without loss of generality we take $\delta m^2_{atm}$ to be
positive.

We do not consider matter effects in our analysis. For the $\delta
m^2_{atm}$ favored by our fits, matter effects are small for the sub-GeV
neutrinos that constitute most of the data~\cite{matter}. Also, as
evidenced by our fits, the dominant oscillation is $\nu_\mu \rightarrow
\nu_\tau$, which is not greatly affected by matter~\cite{pantaleone}.
However, matter effects could be important for the neutrino flux with
smaller $\delta m^2_{atm}/E_\nu$, i.e., multi-GeV data in solutions with
$\delta m^2_{atm}\alt 10^{-3}\rm\,eV^2$, or in long-baseline
experiments~\cite{matter}.

\subsection{Solar experiments}

For neutrinos from the sun $L/E\sim10^{10}$~km/GeV, and the
$\sin^2\Delta_{atm}$ terms oscillate very
rapidly, averaging to ${1\over2}$. Then the
oscillation probabilities are
\bea
P(\nu_e \rightarrow \nu_e) &=& 1 - {1\over2}\sin^22\theta_1
- c_1^4\sin^22\theta_3 \sin^2\Delta_{sun} \,,
\label{Peesun} \\
P(\nu_e \rightarrow \nu_\mu) &=& {1\over2}s_2^2\sin^22\theta_1
+ 4 c_1^2 s_3 c_3 \left[ s_3 c_3 (c_2^2-s_1^2s_2^2)
+ s_1 s_2 c_2 \cos2\theta_3\cos\delta \right] \sin^2\Delta_{sun}
\nonumber
\\
&& - 2 s_1 c_1^2 s_2 c_2 s_3 c_3 \sin\delta \sin2\Delta_{sun} \,,
\label{Pemsun} \\
P(\nu_e \rightarrow \nu_\tau) &=& {1\over2}c_2^2\sin^22\theta_1
+ 4 c_1^2 s_3 c_3 \left[ s_3 c_3 (s_2^2-s_1^2c_2^2)
- s_1 s_2 c_2 \cos2\theta_3\cos\delta \right] \sin^2\Delta_{sun}
\nonumber
\\
&& + 2 s_1 c_1^2 s_2 c_2 s_3 c_3 \sin\delta \sin2\Delta_{sun} \,,
\label{Petsun}
\eea
where $P(\nu_\alpha \rightarrow \nu_\beta) = P(\bar\nu_\beta \rightarrow
\bar\nu_\alpha)$ from $CPT$ invariance and $P(\nu_\beta \rightarrow
\nu_\alpha)$ may be found from $P(\nu_\alpha \rightarrow \nu_\beta)$ by
changing the sign of $\delta$. Since only the sum of oscillation
channels $P(\nu_e \rightarrow \nu_\mu) + P(\nu_e \rightarrow \nu_\tau) =
1 - P(\nu_e \rightarrow \nu_e)$ is tested in solar experiments, the
$CP$-violating parameter $J$ cannot be constrained from solar
measurements. To see the effects of $CP$ violation one must measure
\bea
P(\nu_\mu \rightarrow \nu_e) - P(\nu_e \rightarrow \nu_\mu)
&=& P(\nu_e \rightarrow \nu_\tau) - P(\nu_\tau \rightarrow \nu_e)
\\
&=& P(\nu_\tau \rightarrow \nu_\mu) - P(\nu_\mu \rightarrow \nu_\tau)
\\
&=& 2 s_1 c_1^2 s_2 c_2 s_3 c_3 \sin\delta \sin2\Delta_{sun} \,,
\eea
the corresponding differences for antineutrinos (which have the
opposite sign), or combinations that violate $CP$ explicitly, such as
$P(\nu_\mu \rightarrow \nu_e) - P(\bar\nu_\mu \rightarrow \bar\nu_e)$.

When $\theta_1=0$ (i.e., $U_{e3}=0$), the solar oscillation acts like a
simple two-neutrino oscillation with amplitude $\sin^22\theta_3$;
although the oscillations of $\nu_e$ may involve both $\nu_\mu$ and
$\nu_\tau$, the individual channels $\nu_e \rightarrow \nu_\mu$ and
$\nu_e \rightarrow \nu_\tau$ are not measurable in solar experiments,
since the neutrino energies are below the thresholds for $\mu$ and
$\tau$ production. The parameter $\theta_1$ measures the extent to which
the solar oscillations have a constant component coming from the
atmospheric oscillation scale~\cite{osland}.

In our solar fits we use the shape of the standard solar model (SSM)
neutrino fluxes and absorption cross sections in Ref.~\cite{SSM} and the new
normalizations in Ref.~\cite{BBP}. For the $^{37}$Cl and $^{71}$Ga cases,
we fold the neutrino oscillation probability with the neutrino
absorption cross section and expected neutrino flux to obtain the
expected number of events, which can then be compared to the
expectations of the standard solar model. We also allow an arbitrary
normalization factor $\beta$ for the $^8$B neutrino flux. The expected
number of neutrino events is then
\beq
N = \int \sigma P(\nu_e \rightarrow \nu_e) (\beta \phi_B + \phi_{non-B})
dE_\nu,
\label{N1}
\eeq

For the Super-Kamiokande case, for which the interaction is
$\nu e \rightarrow \nu e$, the events are binned by outgoing electron
energy, and we have included the effects of the detector resolution. The
number of events per unit of electron energy is
\beq
{dN\over dE_e} = \beta \int
\left\{ {d\sigma_{CC}\over dE_e^\prime} P(\nu_e \rightarrow \nu_e)
+ {d\sigma_{NC}\over dE_e^\prime} [1-P(\nu_e \rightarrow \nu_e)] \right\}
G(E_e^\prime, E_e) \phi_B dE_\nu dE_e^\prime
\label{N2}
\eeq
where $d\sigma_{CC}/dE_e$ ($d\sigma_{NC}/dE_e$) is the charge-current
(neutral-current) differential cross section for an incident
neutrino of energy $E_\nu$ and $G(E_e^\prime, E_e)$ is the probability
that an electron of energy $E_e^\prime$ is measured as having energy
$E_e$; the electron energy resolution is taken from
Ref.~\cite{SuperKsolar}. The events are then put into 0.5~GeV bins
starting at the 6.5~GeV threshhold for the detector, to compare with the
Super-Kamiokande data. We take as the input solar data the Homestake
$^{37}$Cl rate~\cite{solar1}, the GALLEX and SAGE~\cite{solar3}
$^{71}$Ga rates, and 16 bins of the Super-Kamiokande detected electron
energy spectrum~\cite{SuperKsolar}. We then fit Eqs.~(\ref{N1}) and
(\ref{N2}) to the data and determine the neutrino parameters in
Eq.~(\ref{Peesun}) and the $^8$B neutrino flux normalization $\beta$.

In our solar fits we combine day and night results, since they should
not be appreciably different for vacuum long-wavelength oscillations.
We compare the model predictions with the time-averaged data, since with
current statistics the Super-Kamiokande data does not reflect any
seasonal variation~\cite{SuperKsolar}. The possibility of detecting a
seasonal variation in Super-Kamiokande and other experiments with
improved statistics is discussed in Sec.~V.

\section{Two-neutrino solutions}

\subsection{Independent solutions for atmospheric and solar data}

In the limit $\theta_1=0$ (i.e., $U_{e3}=0$), the atmospheric and solar
oscillations decouple~\cite{osland,decouple} and effectively reduce to
two separate two-neutrino solutions, each with its own oscillation
amplitude and mass-squared difference. Also, the only oscillation
channel for atmospheric neutrinos is $\nu_\mu \rightarrow \nu_\tau$.
More specifically, for atmospheric neutrinos
\bea
P(\nu_\mu\rightarrow\nu_\mu) &=& 1 - \sin^22\theta_2 \sin^2\Delta_{atm} \,,
\label{pmumu0}\\
P(\nu_e\rightarrow\nu_e) &=& 1 \,,
\label{pee0}\\
P(\nu_e\leftrightarrow\nu_\mu) &=& 0 \,,
\label{pemu0}
\eea
and for solar neutrinos
\beq
P(\nu_e\rightarrow\nu_e) = 1 - \sin^22\theta_3 \sin^2\Delta_{sun} \,.
\label{pee}
\eeq
In the $U_{e3}=0$ limit, fits to the atmospheric and solar data
may be made independently.

\subsection{Atmospheric data}

\begin{figure}
\centering\leavevmode
\epsfysize=6.5in\epsffile{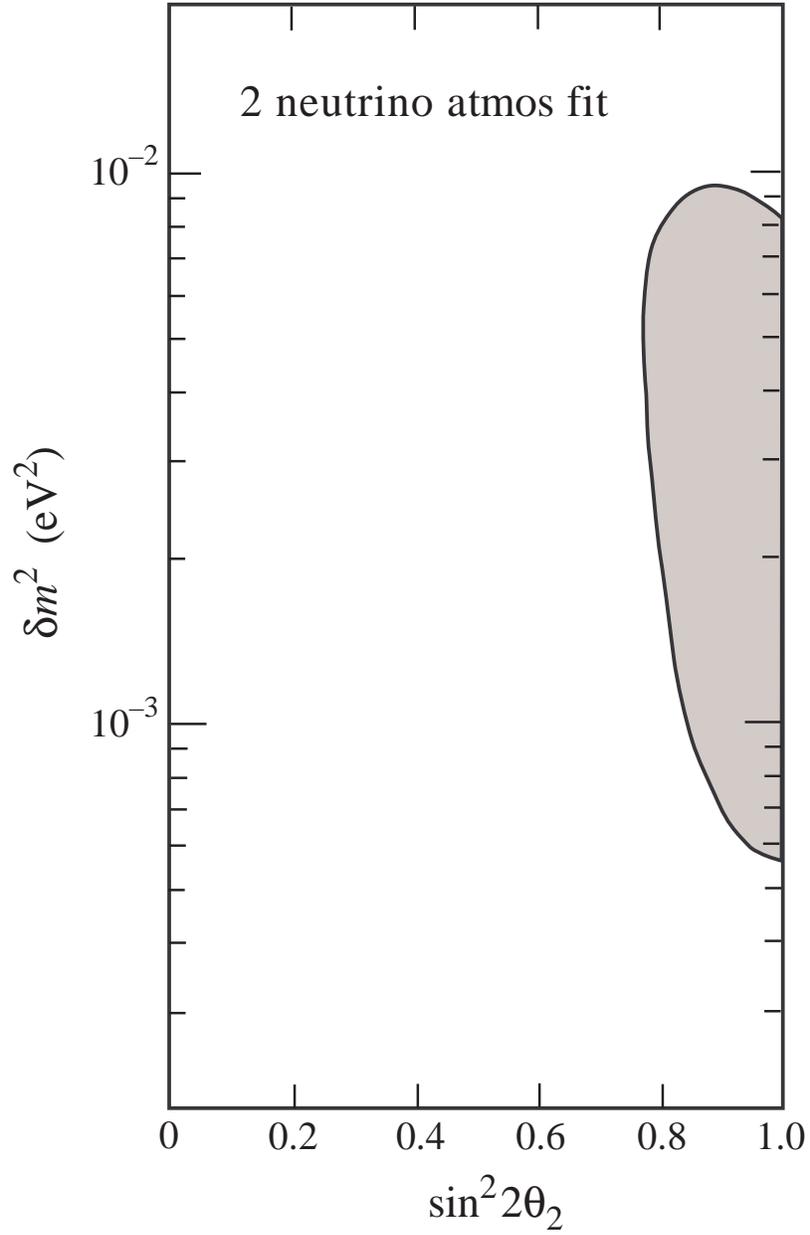}

\caption[]{\label{2nuatm} Allowed region at 95\%~C.L. for our effective
two-neutrino fit to the Super-Kamiokande atmospheric neutrino data.}

\end{figure}

For the two-neutrino case with $\theta_1=0$ (only $\nu_\mu \rightarrow
\nu_\tau$ oscillations) our best fit parameters are
\begin{eqnarray}
\delta m^2_{atm} &=& 2.8\times10^{-3}{\rm~eV}^2 \,,
\label{dm2bestatm}\\
\sin^22\theta_2 &=& 1.00 \,,
\label{ampbestatm}\\
\alpha &=& 1.16 \,,
\label{alphabest}
\end{eqnarray}
where $\alpha$ is the overall flux normalization in Eqs.~(\ref{Rmu}) and
(\ref{Re}), with $\chi^2_{min}=7.1$ for 13 degrees of freedom (DOF).
This $\chi^2/DOF$ corresponds to a goodness-of-fit of 90\%. The
95\%~C.L. allowed region for $\delta m^2_{atm}$ versus $\sin^22\theta_2$
is shown in Fig.~\ref{2nuatm}. Our result is very similar to the fit
obtained by the SuperK collaboration. If we set $\alpha=1$ (i.e., assume
that the theoretical flux normalization is exact), the best fit has
$\chi^2/DOF=22.8/14$, which is acceptable only at the 6\%~C.L. The
calculated flux has a normalization uncertainty of about
$\pm20\%$~\cite{compare}.

As reported by the SuperK collaboration, the other two-neutrino case
with pure vacuum $\nu_\mu \leftrightarrow \nu_e$ oscillations (which
corresponds to $\theta_2=\pi/2$) does not give a good fit to the
data (possible effects of matter are discussed at the end of Sec.~II.B).
We find that the minimum $\chi^2/DOF$ for this case is 81.9/13,
corresponding to a goodness-of-fit of $4\times10^{-12}$. Therefore the
$\nu_\mu \leftrightarrow \nu_e$ scenario for atmospheric neutrinos is
strongly disfavored by the SuperK data. The results of the two-neutrino
fits to the atmospheric data are summarized in Table~\ref{atmfits}. Large
amplitude $\nu_\mu \rightarrow \nu_e$ oscillations are also excluded by
the CHOOZ reactor data~\cite{CHOOZ} for $\delta m^2_{atm} \agt
10^{-3}$~eV$^2$.

\subsection{Solar data}

\begin{figure}
\centering\leavevmode
\epsfysize=6.5in\epsffile{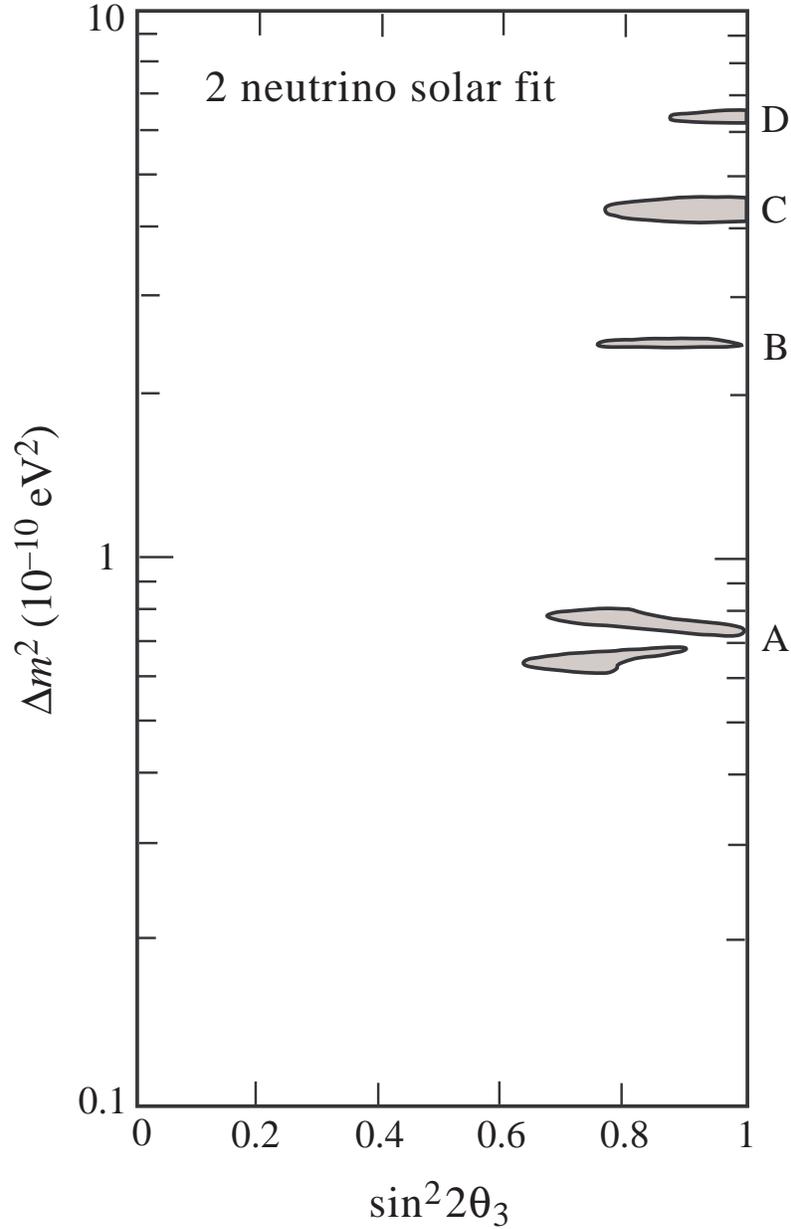}

\caption[]{\label{2nusun} Allowed regions at 95\%~C.L. for our effective
two-neutrino fit to the solar neutrino data from Homestake, SAGE,
GALLEX, and Super-Kamiokande. The four ``finger'' regions from
Table~\ref{sunfits} are labeled A, B, C, and D.}

\end{figure}

For the effective two-neutrino oscillation formula with $\theta_1=0$,
our best fit to the combined solar data yields the parameters
\begin{eqnarray}
\delta m^2_{sun} &=& 7.5\times10^{-11}{\rm~eV}^2 \,,
\label{dm2bestsun}\\
\sin^22\theta_3 &=& 0.91 \,,
\label{ampbestsun}\\
\beta &=& 1.62 \,,
\label{betabest}
\end{eqnarray}
with $\chi^2_{min}/DOF=21.6/16$, acceptable at the 16\%~C.L. The
95\%~C.L. allowed regions for $\delta m^2_{atm}$ versus $\sin^22\theta_3$
are shown in Fig.~\ref{2nusun}. We note that our allowed regions are very
similar to those obtained in Ref.~\cite{bahcall} with a somewhat
different analysis. Taken at face value, the preferred value of the
$^8$B normalization would suggest that the SSM underestimates the $^8$B
neutrino flux by a sizable amount. However, this seems unlikely because
most alternative solar models give a lower $^8$B neutrino flux. The best
fit values for $\beta=1$ (the SSM $^8$B spectrum normalization) are
\begin{eqnarray}
\delta m^2_{sun} &=& 6.5\times10^{-11}{\rm~eV}^2 \,,
\label{dm2bestsun3}\\
\sin^22\theta_3 &=& 0.74 \,,
\label{ampbestsun3}
\end{eqnarray}
with $\chi^2_{min}/DOF=26.5/17$, acceptable at the 7\%~C.L.

We have also searched for the best-fit to
the solar data in each of the $\delta m^2_{sun}$ ``finger'' regions in
Fig.~\ref{2nusun}. The results are given in Table~\ref{sunfits}, where
these four best fits have been labeled A, B, C, D in order of ascending
$\delta m^2_{sun}$. These fits correspond to vacuum oscillation
wavelengths (for a typical $^8$B neutrino energy) of approximately
${1\over2}D$, ${3\over2}D$, ${5\over2}D$, and ${7\over2}D$, where $D$ is
the Earth-Sun distance. We see that the best fits in the higher $\delta
m^2_{sun}$ regions all have a lower $^8$B normalization. They also tend
to fit the Super-K data better, but do somewhat worse on the
radiochemical experiments.  Interestingly, the two solutions with the
largest $\delta m^2_{sun}$ have $\sin^22\theta_3 \simeq 1$.

Because the apparent suppression in the $^{37}$Cl measurement differs
from that of the $^{71}$Ga and SuperK cases, we consider the possibility
that it had an unknown systematic error. The results of fitting to just
the $^{71}$Ga and SuperK data are listed in the lower half of
Table~\ref{sunfits}.  The best fit now occurs for a larger $\delta
m^2_{sun}$, and the goodness-of-fit improves. The allowed regions do
not qualitatively change. However, for the larger $\delta m^2_{sun}$
regions, the $^{71}$Ga data is more easily accomodated by small changes
of the parameters from their values in the global fit.

\subsection{Existence of separate mass scales}

In making our three-neutrino fits we assume that separate mass-squared
difference scales are necessary to explain the atmospheric and solar
neutrino data. If the $\delta m^2$ scales were not distinct, or if one
of the mass-squared differences were used to explain the LSND data, then
either the solar or atmospheric probabilities would be in a region of
$L/E$ where the oscillations have averaged, and there would be no energy
dependence.  An energy-independent suppression due to oscillations
would be equivalent to letting the overall normalization vary.
Oscillation scenarios where there is not a separate mass scale
associated with solar neutrinos have been considered~\cite{conforto}.

It has already been demonstrated in the literature that an energy-independent
suppression is strongly disfavored by the solar data~\cite{flatsol}.
Even if one ignores the $^{37}$Cl data, and
assumes that the $^{71}$Ga and $\nu e$ experimental rates are both
consistent with an overall suppression by 50\%, the spectrum distortion
of $^8$B neutrinos measured in the $\nu e$ experiments disfavors an
energy-independent suppression. We have updated this analysis, allowing
for an overall flux suppression in addition to the variation of the
$^8$B neutrino normalization. The lowest value $P(\nu_e \rightarrow
\nu_e)$ can achieve with three neutrinos when all oscillations are
averaged is $1\over3$. We found the best fit to the solar data with
overall depletion between $1\over3$ and unity and arbitrary $^8$B
normalization has $\chi^2/DOF=48.1/17$, which is ruled out at the
99.99\%~C.L. If the $^{37}$Cl data are ignored, the best fit has only
$\chi^2/DOF=25.0/16$, ruled out at the 93\%~C.L. Therefore the
distortion of the electron energy spectrum present in the solar neutrino
data favors the existence of a separate mass scale for the oscillation
of the solar neutrinos, which justifies the form of
Eqs.~(\ref{pmumu2})--(\ref{pmutau2}) and (\ref{Peesun})--(\ref{Petsun}).

If one $\delta m^2$ is used to describe the LSND data and the other is
used to describe the solar data, then the oscillation due to $\delta
m^2_{LSND}$ will be averaged for the $L/E$ of the atmospheric neutrinos,
so that the oscillation probabilities are independent of both energy and
zenith angle. We find the best fit in this scenario to have $\chi^2/DOF
= 33.2/13$, which is excluded at the 99.8\%~C.L.

\section{Three-neutrino solutions}

\subsection{Bi-maximal solution}

The atmospheric data favor maximal mixing of atmospheric $\nu_\mu$ with
$\nu_\tau$ and no mixing with $\nu_e$. The solar data also suggest,
although not as strongly, that solar neutrinos may also mix maximally,
or nearly maximally. If we require both atmospheric
and solar oscillations to be maximal, there is a unique three-neutrino
solution to the neutrino mixing matrix~\cite{bimax}, which corresponds to
$\theta_1=0$ and $\theta_2=\theta_3=\pi/4$. The corresponding
oscillation probabilities for atmospheric neutrinos are
\bea
P(\nu_\mu\rightarrow\nu_\mu) &=& 1 - \sin^2\Delta_{atm} \,,
\label{pmumubi}\\
P(\nu_e\rightarrow\nu_e) &=& 1 \,,
\label{peebi}\\
P(\nu_e\leftrightarrow\nu_\mu) &=& 0 \,,
\label{pemubi}
\eea
and for solar neutrinos
\bea
P(\nu_e\rightarrow\nu_e) &=& 1 - \sin^2\Delta_{sun} \,.
\label{peebisun} \,,
\\
P(\nu_e \rightarrow \nu_\mu) = P(\nu_e \rightarrow \nu_\tau) &=&
{1\over 2} \sin^2\Delta_{sun}
\label{pembisun} \,.
\eea
One interesting aspect of this solution is that the solar $\nu_e$
oscillations are 50\% into $\nu_\mu$ and 50\% into $\nu_\tau$, although
the flavor content of the $\nu_e$ oscillation is not observable in solar
experiments. Further properties of the bi-maximal and nearly
bi-maximal solutions are discussed in Ref.~\cite{bimax}.

\subsection{Atmospheric data}

\begin{figure}
\centering\leavevmode
\epsfysize=6.5in\epsffile{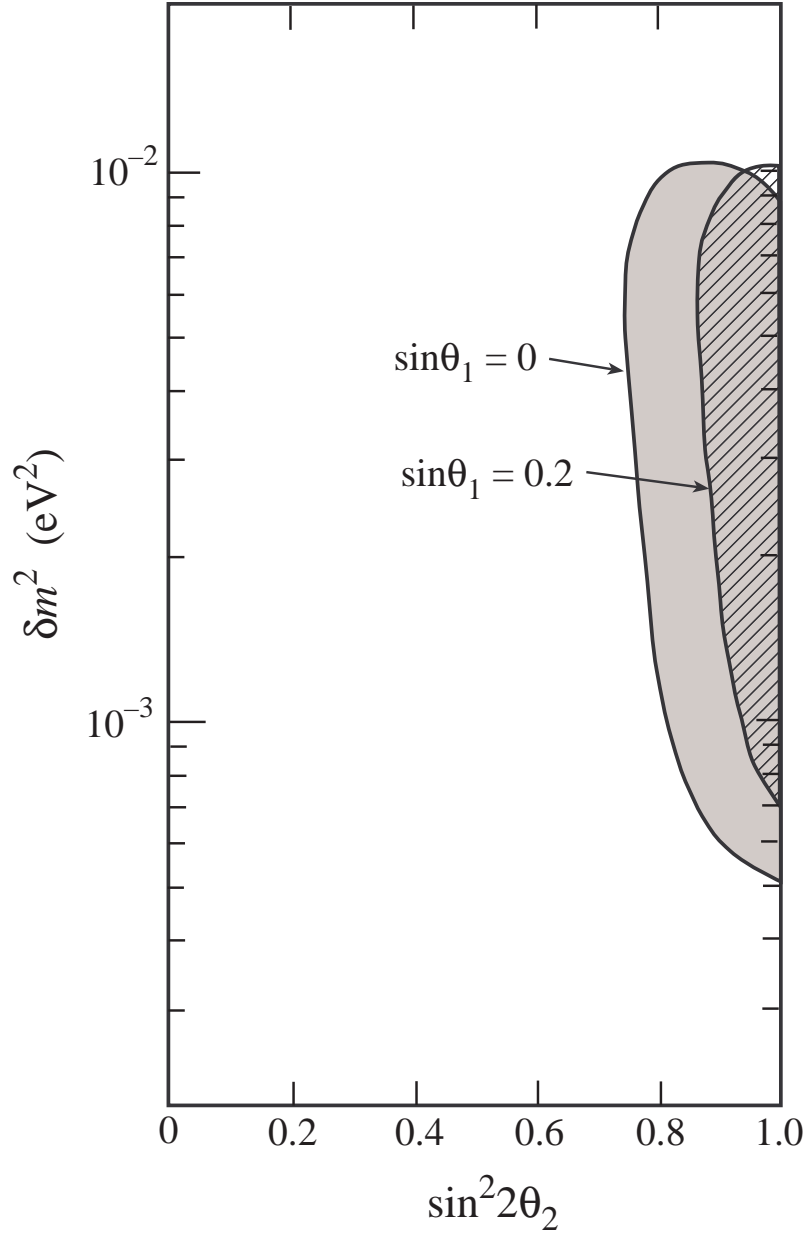}

\caption[]{\label{3nuatm} Allowed regions at 95\%~C.L. from the
Super-Kamiokande atmospheric neutrino data for $\delta
m^2_{atm}$ versus $\sin^22\theta_2$ when the overall atmospheric neutrino
flux normalization $\alpha$ is allowed to vary, for $\sin\theta_1=0$
and $\sin\theta_1=0.2$.}

\end{figure}

\begin{figure}
\centering\leavevmode
\epsfysize=6in\epsffile{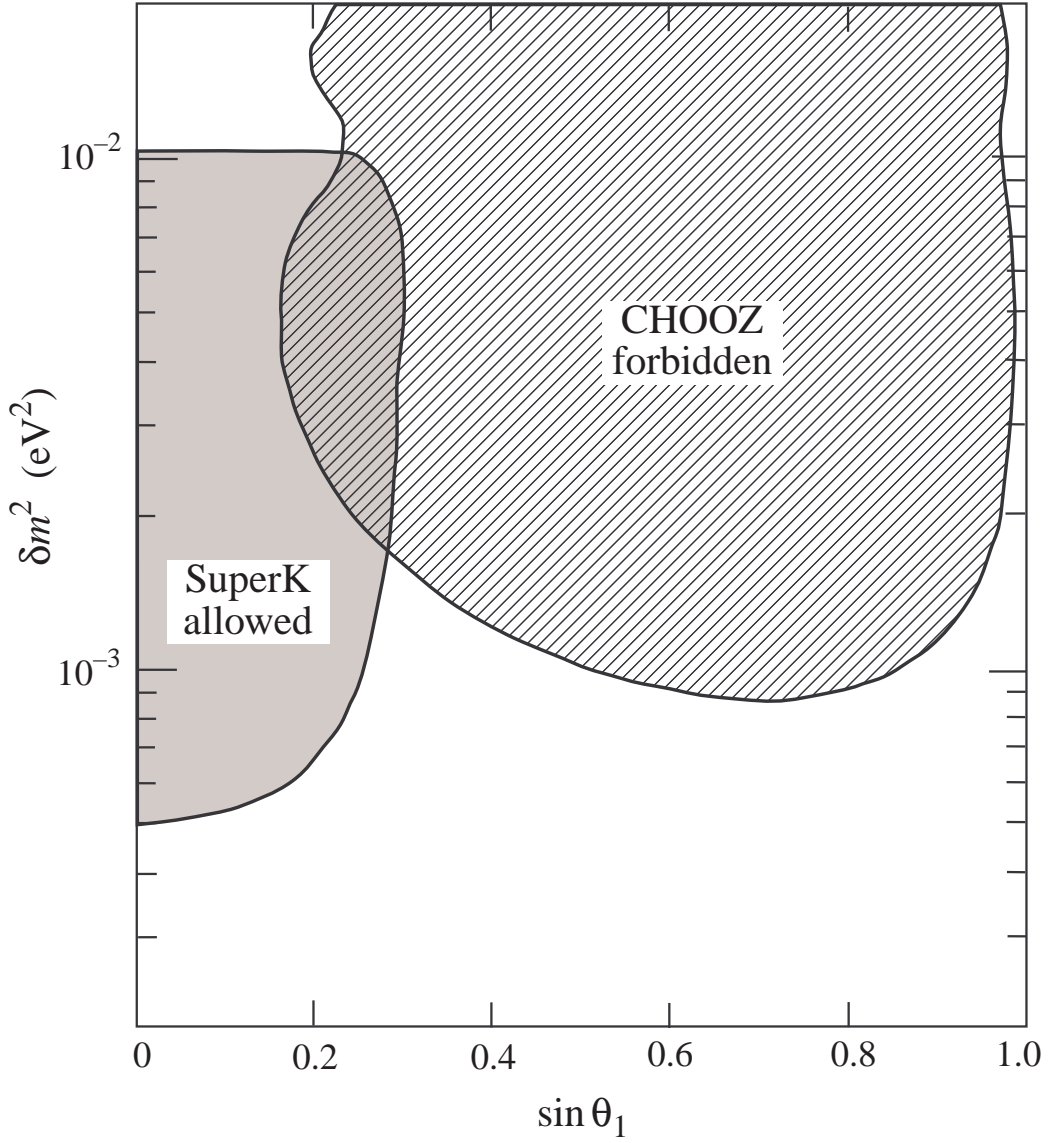}

\caption[]{\label{3nuatm2} Allowed regions at 95\%~C.L. from the
Super-Kamiokande atmospheric neutrino data for $\delta
m^2_{atm}$ versus $\sin\theta_1$ when $\sin^22\theta_2$ and the overall
atmospheric neutrino flux normalization $\alpha$ are allowed to vary.
The CHOOZ constraint is also shown.}

\end{figure}

A full three-neutrino fit to the atmospheric data with one $\delta
m^2_{atm}$ scale has been made in Refs.~\cite{bww98,3nuatm}. In terms of
the oscillation parameters defined in Sec.~II, our best fit values for
the four parameters are
\begin{eqnarray}
\delta m^2_{atm} &=& 2.8\times10^{-3}{\rm~eV}^2 \,,
\label{dm2bestatm2}\\
\sin\theta_1 &=& 0.00 \,,
\label{t1bestatm}\\
\sin^22\theta_2 &=& 1.00 \,,
\label{ampbestatm2}\\
\alpha &=& 1.16 \,,
\label{alphabest2}
\end{eqnarray}
with $\chi^2_{min}/DOF=7.1/12$, acceptable at the 85\%~C.L. 
In Fig.~\ref{3nuatm} we show the
95\%~C.L. allowed region for $\sin^22\theta_2$ versus $\delta m^2_{atm}$
for various values of $\sin\theta_1$ when the flux normalization
$\alpha$ is allowed to vary. Although $\sin\theta_1=0$ is favored,
nonzero values are allowed, which permit some $\nu_\mu \leftrightarrow
\nu_e$ and $\nu_e \rightarrow \nu_\tau$ oscillations of atmospheric
neutrinos. In Fig.~\ref{3nuatm2} we show the 95\%~C.L. allowed region
for $\sin\theta_1$ versus $\delta m^2_{atm}$ when $\alpha$ and
$\sin^22\theta_2$ are allowed to vary.

Another limit on $\sin\theta_1$ comes from the CHOOZ reactor
experiment~\cite{CHOOZ} that measures $\bar\nu_e$ disappearance
\beq
A^{e\not e}_{atm} = 4 P_{e3} (1 - P_{e3}) \alt 0.2 \,,
\label{choozlimit}
\eeq
which applies for $\delta m^2_{atm} \agt 2\times10^{-3}$~eV$^2$. The
exact limit on $A^{e\not e}_{atm}$ varies with $\delta m^2_{atm}$, and
for $\delta m^2_{atm} < 10^{-3}$~eV$^2$ there is no limit at all. For
$\delta m^2_{atm} = 2.8\times10^{-3}$~eV$^2$ and $\alpha=1.16$,
$\sin\theta_1$ is constrained to be less than 0.19. The result of
imposing the CHOOZ constraint is shown in Figs.~\ref{3nuatm} and
\ref{3nuatm2}. In Fig.~\ref{3nuatm2} we see that the range of
$\sin\theta_1$ allowed by the fit to the atmospheric neutrino data and
the CHOOZ constraint is
\beq
0 \le \sin\theta_1 \le 0.29 \,,
\label{sinth1atm}
\eeq
at 95\%~C.L.

\subsection{Solar data}

\begin{figure}
\centering\leavevmode
\epsfxsize=6.5in\epsffile{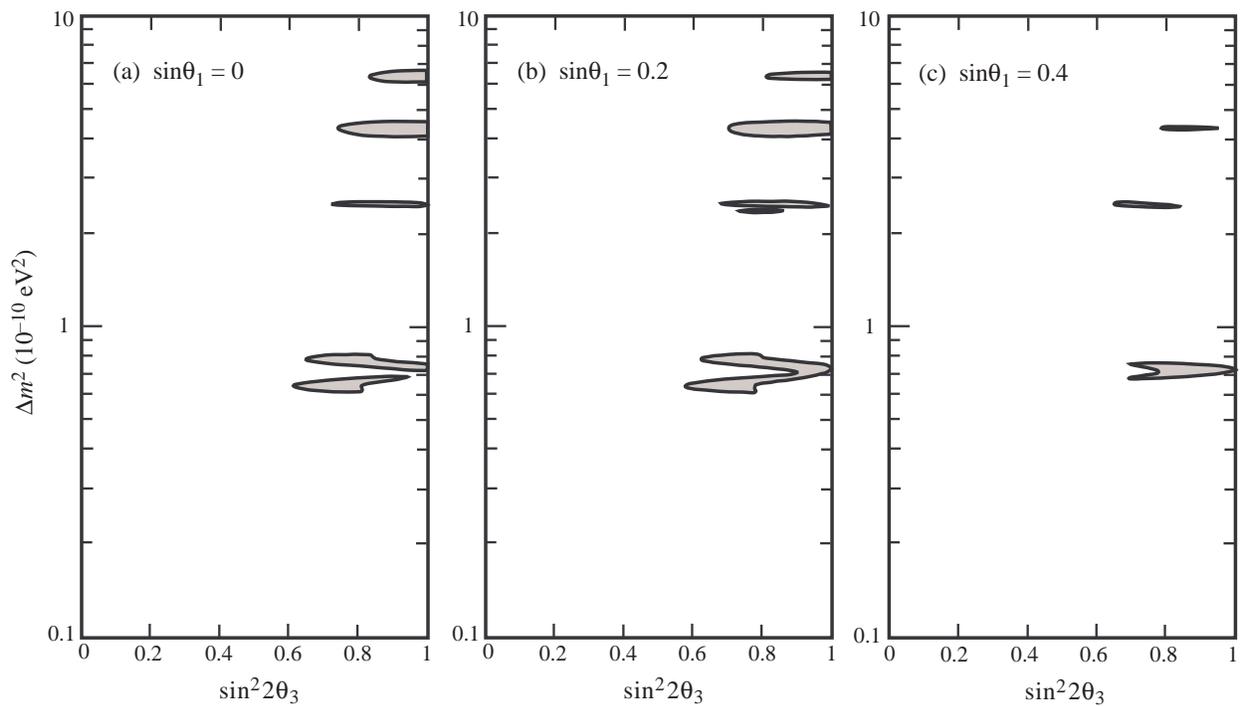}

\caption[]{\label{3nusun} Allowed regions at 95\%~C.L. from the solar
neutrino data for $\delta m^2_{sun}$ versus $\sin^22\theta_3$ when the
$^8$B neutrino flux normalization $\beta$ is allowed to vary, for (a)
$\sin\theta_1=0$, (b) $\sin\theta_1=0.2$, and (c) $\sin\theta_1=0.4$}

\end{figure}

\begin{figure}
\centering\leavevmode
\epsfysize=6.5in\epsffile{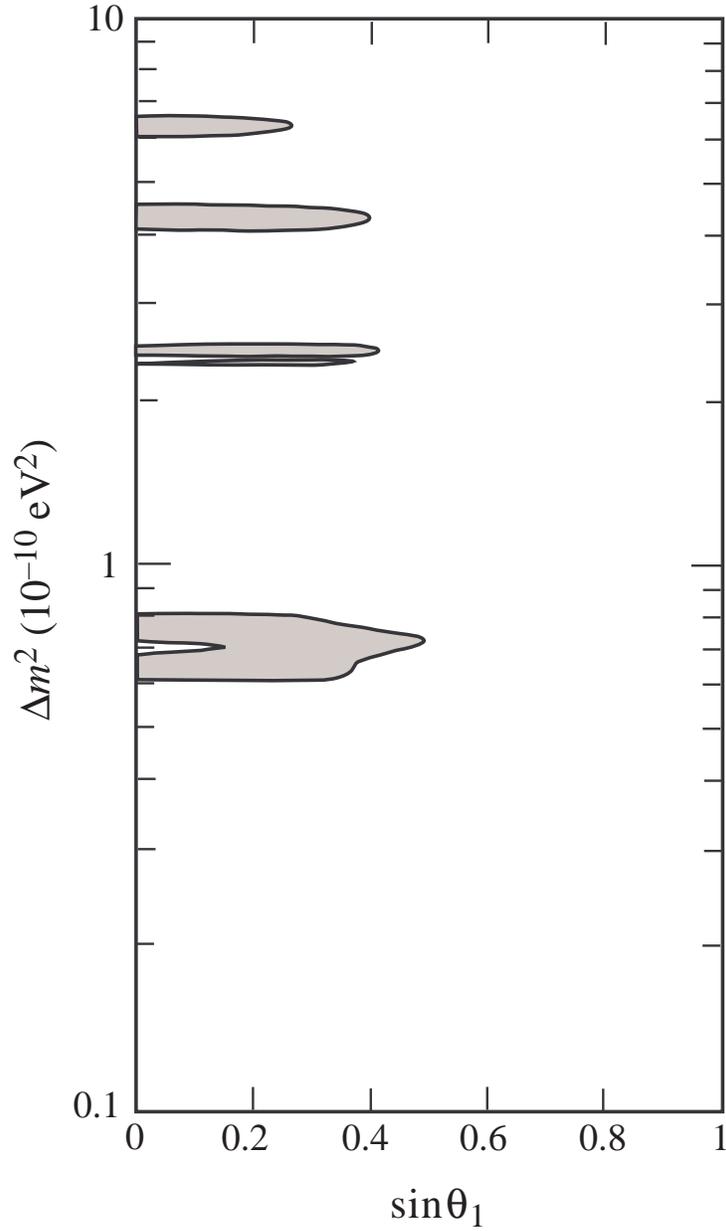}

\caption[]{\label{3nusun2} Allowed regions at 95\%~C.L. from the
solar neutrino data for $\delta
m^2_{sun}$ versus $\sin\theta_1$ when $\sin^22\theta_3$ and the $^8$B
neutrino flux normalization $\beta$ are allowed to vary.}

\end{figure}

A full three-neutrino fit to solar data can be made by varying
$\theta_1$, $\theta_3$, $\delta m^2_{sun}$ and the $^8$B flux normalization
$\beta$, and using the expression in Eq.~(\ref{Peesun}) for the
oscillation probabilities in Eqs.~(\ref{N1}) and (\ref{N2}). We find the
best fit values for the solar parameters
\begin{eqnarray}
\delta m^2_{atm} &=& 7.5\times10^{-11}{\rm~eV}^2 \,,
\label{dm2bestsun2}\\
\sin\theta_1 &=& 0.00 \,,
\label{t1bestsun}\\
\sin^22\theta_3 &=& 0.91 \,,
\label{ampbestsun2}\\
\beta &=& 1.62 \,,
\label{betabest2}
\end{eqnarray}
with $\chi^2_{min}/DOF=21.6/15$, acceptable at the 12\%~C.L. If the
$^8$B normalization is fixed at the SSM value ($\beta=1$), the best fit
is
\begin{eqnarray}
\delta m^2_{atm} &=& 6.5\times10^{-11}{\rm~eV}^2 \,,
\label{dm2bestsun4}\\
\sin\theta_1 &=& 0.00 \,,
\label{t1bestsun4}\\
\sin^22\theta_3 &=& 0.74 \,,
\label{ampbestsun4}
\end{eqnarray}
with $\chi^2_{min}/DOF=26.5/16$, acceptable at the 5\%~C.L. The addition
of the extra parameter $\sin\theta_1$ does not improve the fit to the
solar data. In Fig.~\ref{3nusun} we show the 95\%~C.L. allowed region
for $\sin^22\theta_3$ versus $\delta m^2_{sun}$ for various values of
$\sin\theta_1$ when $\beta$ is allowed to vary. As in the atmospheric
neutrino case, $\sin\theta_1=0$ is favored, although non-negligible
nonzero values are allowed. In Fig.~\ref{3nusun2} we show the
95\%~C.L. allowed region for $\sin\theta_1$ versus $\delta m^2_{sun}$
when $\beta$ and $\sin^22\theta_3$ are allowed to vary. The range of
$\sin\theta_1$ allowed by the solar data is
\beq
0 \le \sin\theta_1 \le 0.49 \,,
\label{sinth1sun}
\eeq
at 95\%~C.L.

\subsection{Global atmospheric and solar fit}

\begin{figure}
\centering\leavevmode
\epsfysize=6.5in\epsffile{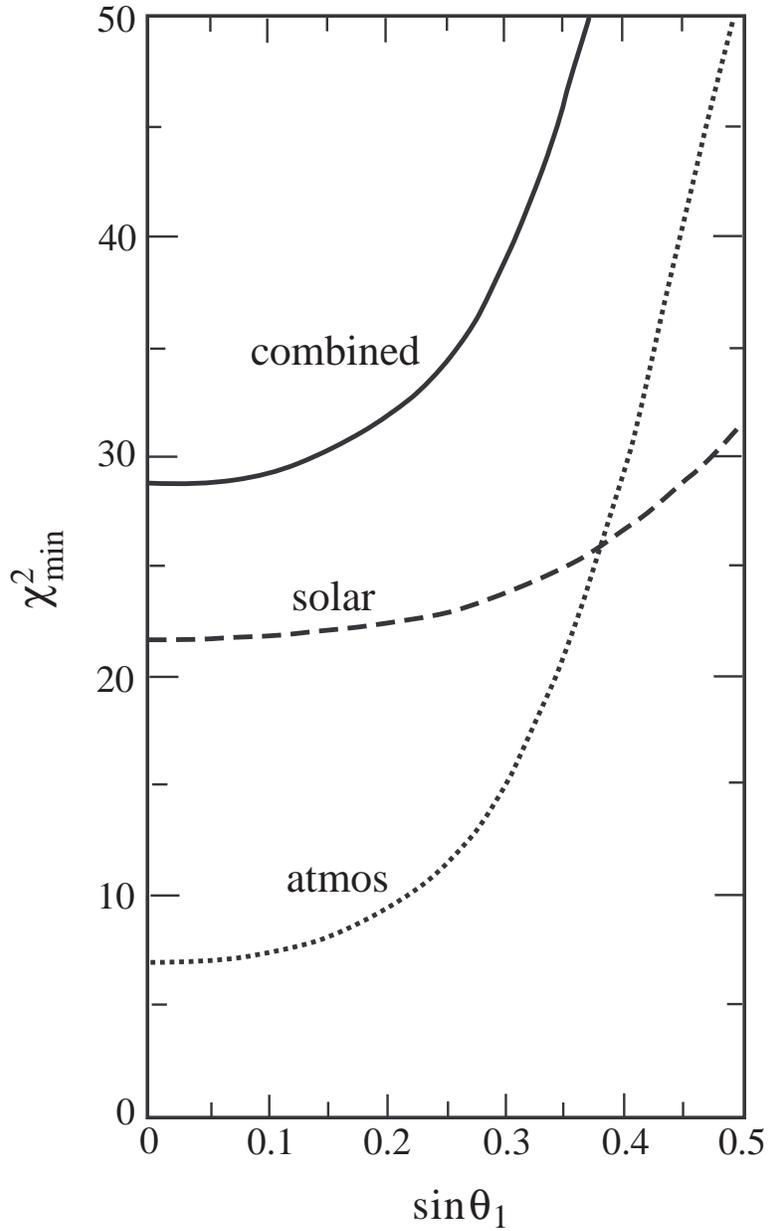}

\caption[]{\label{chi2} Minimum $\chi^2$ versus $\sin\theta_1$ for
the atmsopheric data set (dotted line), the solar data set (dashed), and
the combined atmospheric and solar data sets (solid). All other relevant
parameters are allowed to vary in these fits.}

\end{figure}

Since the atmospheric and solar fits have only one common parameter,
$\sin\theta_1$, and the best separate fits to the two data sets both
have $\sin\theta_1=0$, then the best fit to the combined solar and
atmospheric data sets can be identified immediately as given by
Eqs.~(\ref{dm2bestatm2})-(\ref{alphabest2}) and
(\ref{dm2bestsun2})-(\ref{betabest2}), with $\chi^2_{min}/DOF=28.7/28$,
acceptable at the 43\%~C.L. To see how the fits vary with the common
parameter, in Fig.~\ref{chi2} we show $\chi^2_{min}$ for the solar and
atmospheric data sets versus $\sin\theta_1$, as well as the combined
$\chi^2$. From the figure we see that the dependence of $\chi^2_{min}$
on $\sin\theta_1$ is relatively weak for $\sin\theta_1<0.3$ for the
solar data set. Therefore, the parameter regions allowed by the combined
solar and atmospheric data sets are essentially the most stringent of
those allowed by the solar and atmospheric data sets separately.

\section{Discussion}

\subsection{Long-baseline oscillations}

The MINOS~\cite{MINOS}, K2K~\cite{K2K}, ICARUS~\cite{ICARUS} and
NOE~\cite{ICARUS} experiments can test for $\nu_\mu
\rightarrow \nu_x$ oscillations for $\delta m^2_{atm} > 10^{-3}$~eV$^2$,
and MINOS and ICARUS can also search for $\nu_\mu \rightarrow \nu_\tau$.
Together these measurements could more precisely determine
$\sin^22\theta_2$, $\sin\theta_1$, and $\delta m^2_{atm}$. Further
measurements of atmospheric neutrinos will also help constrain these
parameters. Full three-neutrino fits including the solar neutrino
data~\cite{osland} can then determine one of the remaining two
independent parameters in the mixing matrix, e.g., $\sin^22\theta_3$,
and $\delta m^2_{sun}$.

\begin{figure}
\centering\leavevmode
\epsfysize=6.5in\epsffile{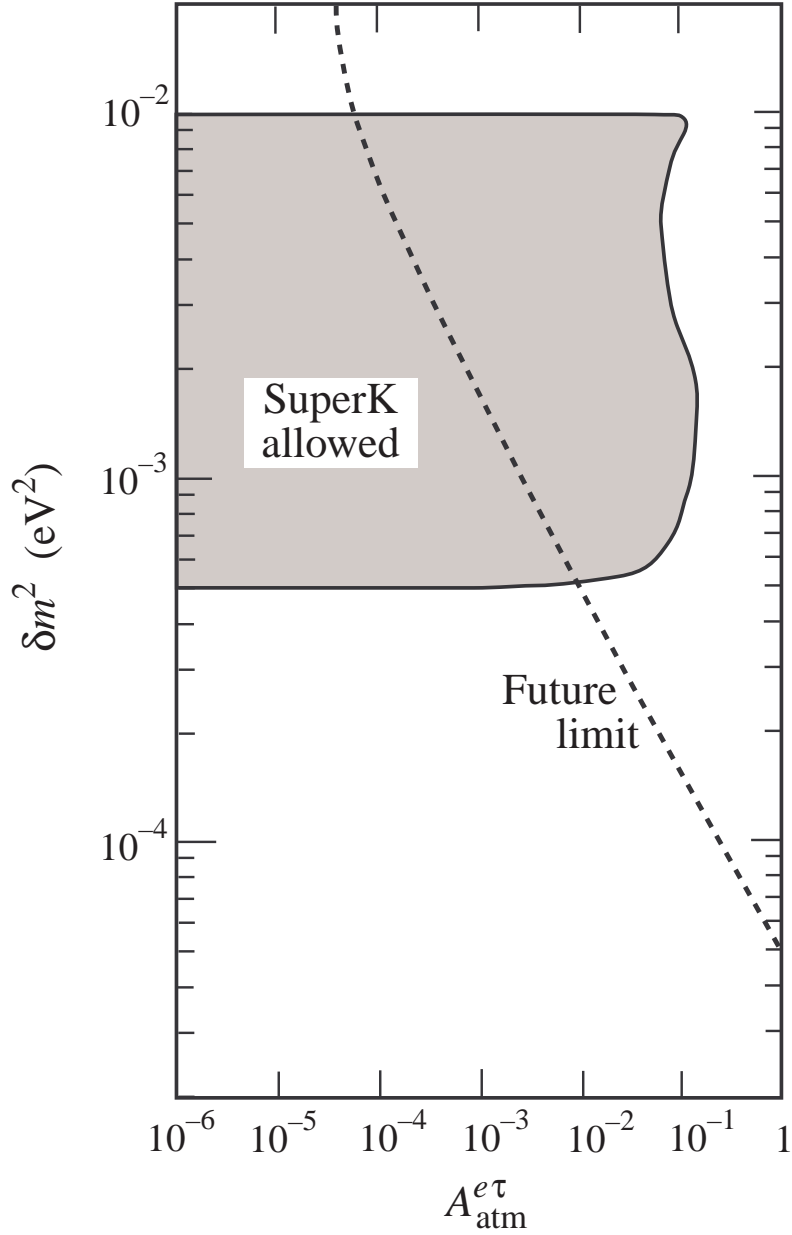}

\caption[]{\label{etau} Allowed region at 95\%~C.L. of the $\nu_e \rightarrow
\nu_\tau$ oscillation amplitude in atmospheric and long-baseline
experiments versus $\delta m^2_{atm}$, determined from a fit to the
atmospheric neutrino data set when the CHOOZ constraint is included.
The expected sensitivity of a long-baseline experiment from Fermilab to
Soudan using muon storage rings is shown by the dashed line.}

\end{figure}

There is new physics predicted when $\sin\theta_1 \ne 0$, i.e.,
$\nu_e\rightarrow\nu_\tau$ oscillations with leading probability given
by Eq.~(\ref{petau2}). The allowed range at 95\%~C.L. for the $\nu_e
\rightarrow \nu_\tau$ oscillation amplitude $A^{e\tau}_{atm}$ versus
$\delta m^2_{atm}$ is depicted in Fig.~\ref{etau}; the effect of the
CHOOZ constraint is also shown. The maximal $\nu_e \rightarrow \nu_\tau$
amplitude of about 0.15
occurs at $\delta m^2_{atm} = 1.7\times10^{-3}$~eV$^2$. These
$\nu_e\rightarrow\nu_\tau$ oscillations could be observed by
long-baseline neutrino experiments with proposed high intensity muon
sources~\cite{geer,4nu,derujula}, which can also make precise
measurements of $\nu_\mu \leftrightarrow \nu_e$ and $\nu_\mu \rightarrow
\nu_\tau$ oscillations. Sensitivity to $A^{e\tau}_{atm} (\delta
m^2_{atm} / {\rm eV}^2)^2 > 2.5\times10^{-9}$ is expected~\cite{geer}
for the parameter ranges of interest here; matter effects would not be
large for an experiment from Fermilab to Soudan~\cite{matter}. For
$\delta m^2_{atm} =
2.8\times10^{-3}$~eV$^2$, $A^{e\tau}_{atm}$ could be measured down to
$3\times10^{-4}$; the expected sensitivity versus $\delta m^2_{atm}$ is
shown in Fig.~\ref{etau}. Precise measurement of
$\nu_e\rightarrow\nu_\tau$ and $\nu_\mu \leftrightarrow \nu_e$
oscillations in such a long-baseline experiment would uniquely specify
the $CP$-conserving part of the three-neutrino model.

\subsection{Future solar tests}

In this section we discuss how future solar neutrino measurements can
distinguish between the different solar vacuum oscillation solutions.
We consider only the two-neutrino solutions since the discussion easily
generalizes to the three-neutrino case.
 
\begin{figure}
\centering\leavevmode
\epsfxsize=6in\epsffile{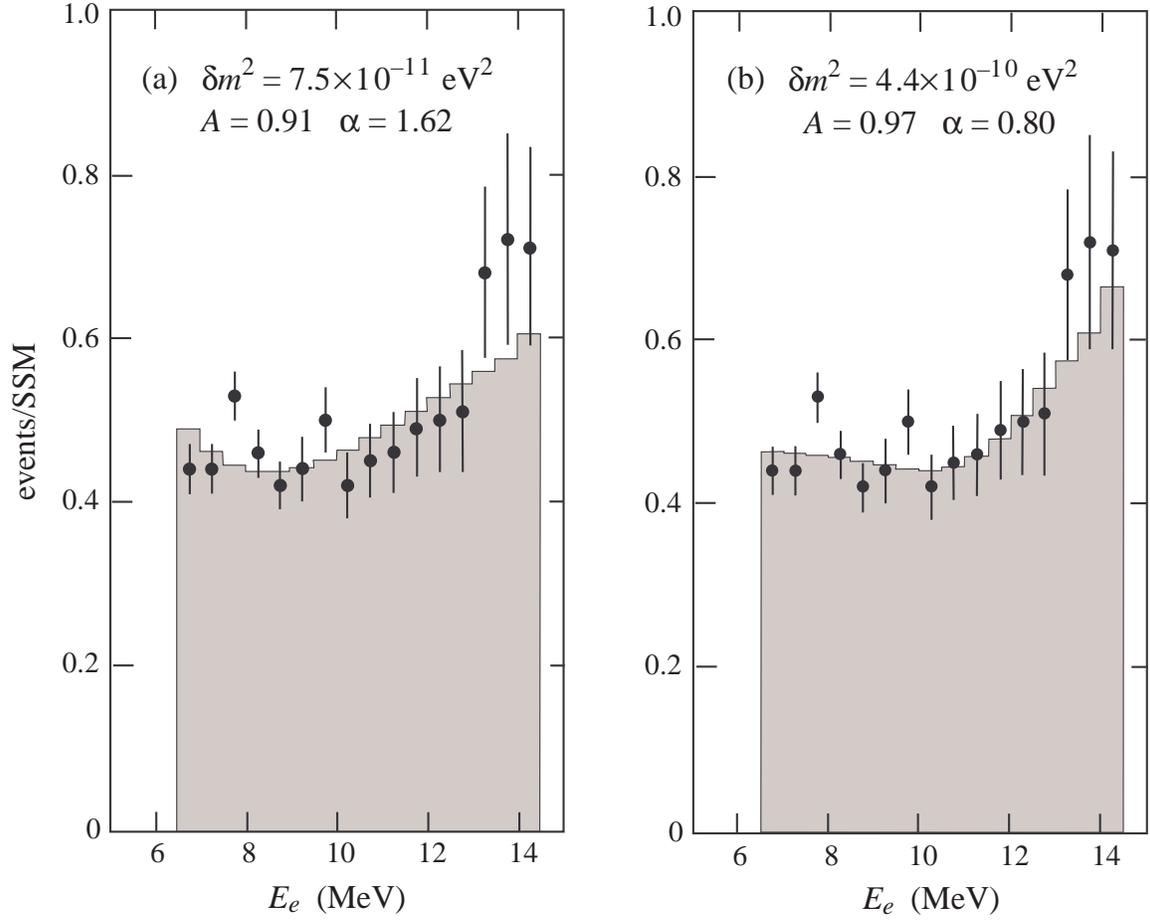}

\caption[]{\label{spectrum} Ratio of the electron energy spectrum to the
SSM prediction for two different two-neutrino vacuum long-wavelength
oscillation scenarios, compared to the Super-Kamiokande data. The
shaded histograms show the results when oscillations are included. The
two cases shown are (a) the best overall fit to all solar neutrino data,
solution A from Table~\ref{sunfits}, and (b) the best fit to just the
Super-Kamiokande data, solution C from Table~\ref{sunfits}.}

\end{figure}

\begin{figure}
\centering\leavevmode
\epsfysize=6.5in\epsffile{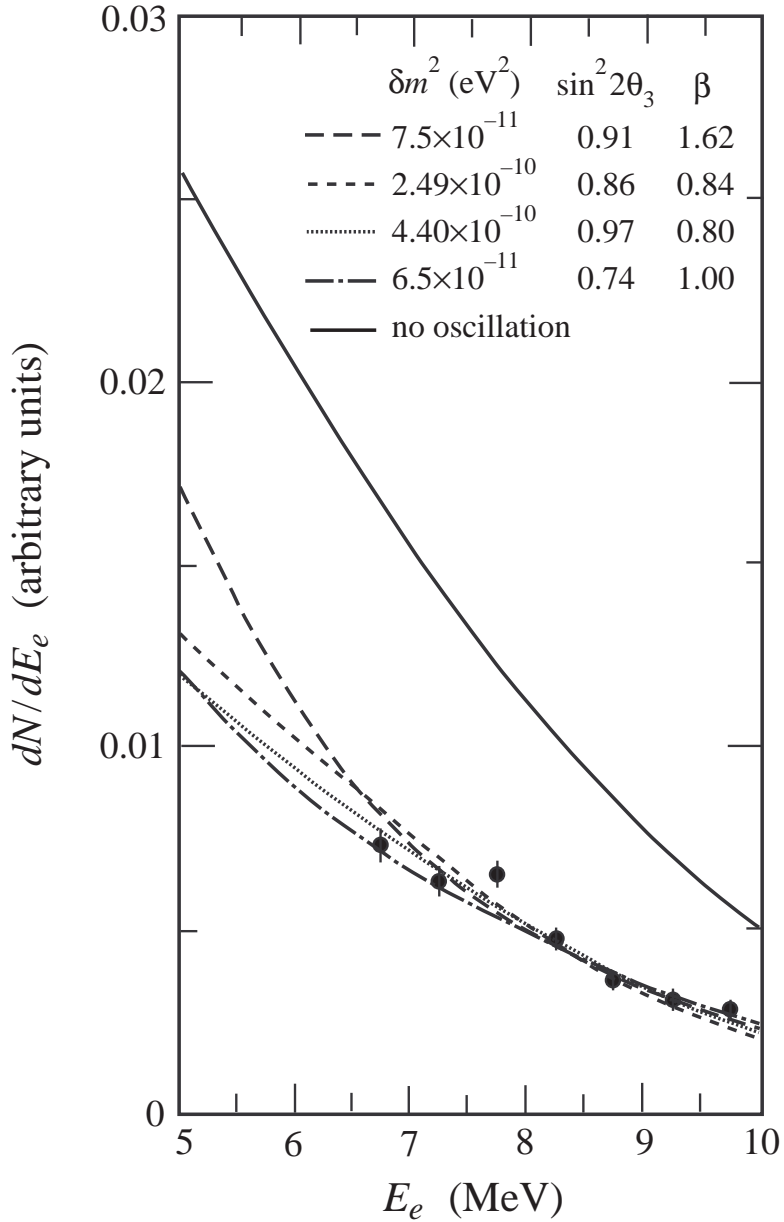}

\caption[]{\label{lowE} Electron energy spectrum in
Super-Kamiokande for the Standard Solar Model with no
oscillations (solid curve) and for the four vacuum neutrino
oscillation solutions A (long-dashed), B (short-dashed), C (dotted),
and D (dash-dotted) listed in Table~\ref{sunfits}. Also shown is the
current data from Super-Kamiokande~\cite{SuperKsolar}.}

\end{figure}

One interesting feature of vacuum long-wavelength solutions is that
they can give a rise in the fraction of surviving $\nu_e$'s for higher
electron energies, in agreement with the SuperK
measurement~\cite{SuperKsolar}.  Figure~\ref{spectrum} shows the ratio
of the electron energy spectrum to the SSM prediction for two different
vacuum long-wavelength scenarios and the SuperK data. Future
measurements at SuperK will improve the statistics in the high $E_e$
bins. However, with an $hep$ flux that is $\sim25$ times greater than
the SSM result, this rise at high $E_e$ could be due to contributions of
$hep$ neutrinos~\cite{hep}. SuperK is also planning to lower their
threshhold for electron detection to 5~MeV; this is particularly
important since the number of events rises for energies just below the
current threshhold of 6.5~MeV. In Fig.~\ref{lowE} we show predictions
for the electron energy spectrum in SuperK in the range 5~MeV~$\le E_e
\le$~10~MeV for the four oscillation solutions A, B, C, and D in
Table~\ref{sunfits}. It is evident that Solutions A and B may be
distinguishable from C and D, and from each other, using data at lower
$E_e$. The energy spectrum measured in the SNO experiment~\cite{SNO}
will also provide additional information on the energy dependence of the
spectrum suppression.

Another feature of vacuum solar solutions is that they may cause a
detectable seasonal variation as the distance between the Earth and Sun
varies~\cite{seasonal1,seasonal2}. We define two seasonal asymmetry
parameters,
\bea
A_1 &=& {2(N_W - N_S) \over (N_W + N_S + N_F + N_{SP})} \,,
\label{asy1}
\\ A_2 &=& {N_W + N_S - N_F - N_{SP} \over N_W + N_S + N_F + N_{SP}} \,,
\label{asy2}
\eea
where $N_W$, $N_{SP}$, $N_S$, and $N_F$ are the number of events
collected in the time periods from November 20 to February 19 (winter,
Earth closest to the Sun), February 20 to May 21 (spring), May 22 to
August 20 (summer, Earth farthest from the Sun), and August 21 to
November 19 (fall), respectively. Since $N_F = N_{SP}$ (the same range
of distances is covered), $A_1$ and $A_2$ are the only independent
quantities that may be constructed from the four seasonal
measurements. The quantity $A_1$ is similar to the seasonal asymmetry
defined in Ref.~\cite{MS}, and $A_1$ and $A_2$ are similar to the first
two harmonics in the analysis of the first paper in
Ref.~\cite{seasonal2}. The parameter $A_1$ has a significant nonzero
value when the oscillation probability increases or decreases
monotonically as the Earth moves from perihelion to aphelion, while
$A_2$ has a significant nonzero value when the oscillation probability
reaches a local extremum somewhere between perihelion and aphelion. Due
to the $1/r^2$ dependence of the solar neutrino flux and the 3.3\%
change in $r$ from perihelion to aphelion, $A_1$ and $A_2$ have the
values 0.030 and 0, respectively, in the absence of oscillations. The
asymmetry $A_1$ is strongly correlated with the electron energy spectrum
distortion; such a correlation is a distinctive characteristic of vacuum
oscillations~\cite{MS}.

\begin{figure}
\centering\leavevmode
\epsfysize=6.5in\epsffile{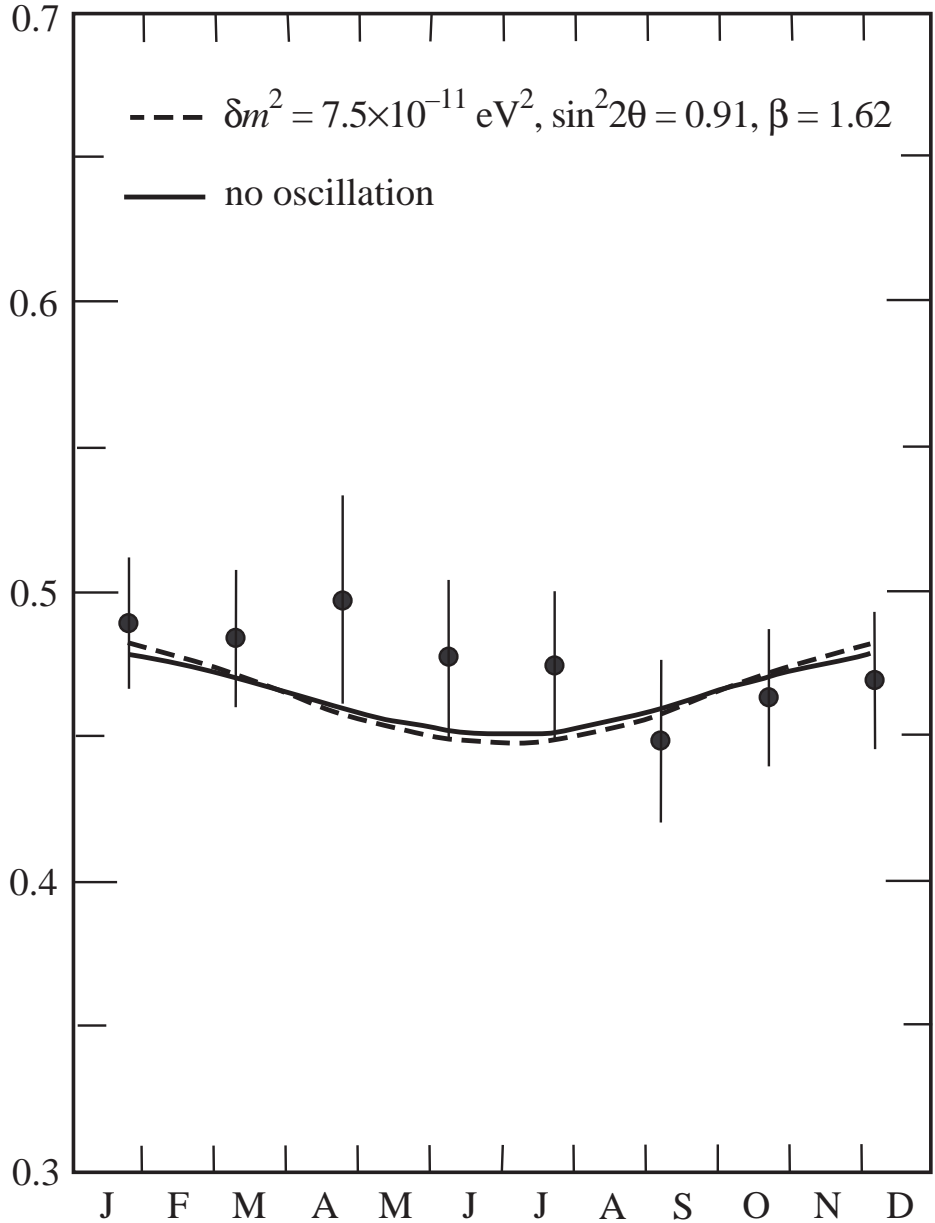}

\caption[]{\label{SKseas} Predicted event rate as a fraction of the SSM
value in Super-Kamiokande versus time of the year for vacuum
oscillation Solution~A in Table~\ref{sunfits} (dashed curve) and for no
oscillations when the overall event rate is normalized to the observed
value (solid). The other solutions in Table~\ref{sunfits} give
similar results. Also shown is the current data from
Super-Kamiokande~\cite{SuperKsolar}.}

\end{figure}

For $^8$B neutrinos, due to their relatively higher energies and longer
oscillation wavelengths, the seasonal variation is less pronounced than
for solar neutrinos of lower energy. The total SuperK event rate as a
fraction of the SSM value versus time of year~\cite{SuperKsolar} is
shown in Fig.~\ref{SKseas}; also shown is the prediction for Solution~A
in Table~\ref{sunfits} and the prediction for no oscillations. The
results for other oscillation solutions are similar to the prediction
for Solution~A. Oscillation solutions produce an enhanced seasonal
effect~\cite{seasonal1,seasonal2,MS}. The SuperK experiment has not
observed a significant seasonal variation with the current data sample,
but could be sensitive to the effects predicted by vacuum oscillations
with increased statistics.

\begin{figure}
\centering\leavevmode
\epsfxsize=6in\epsffile{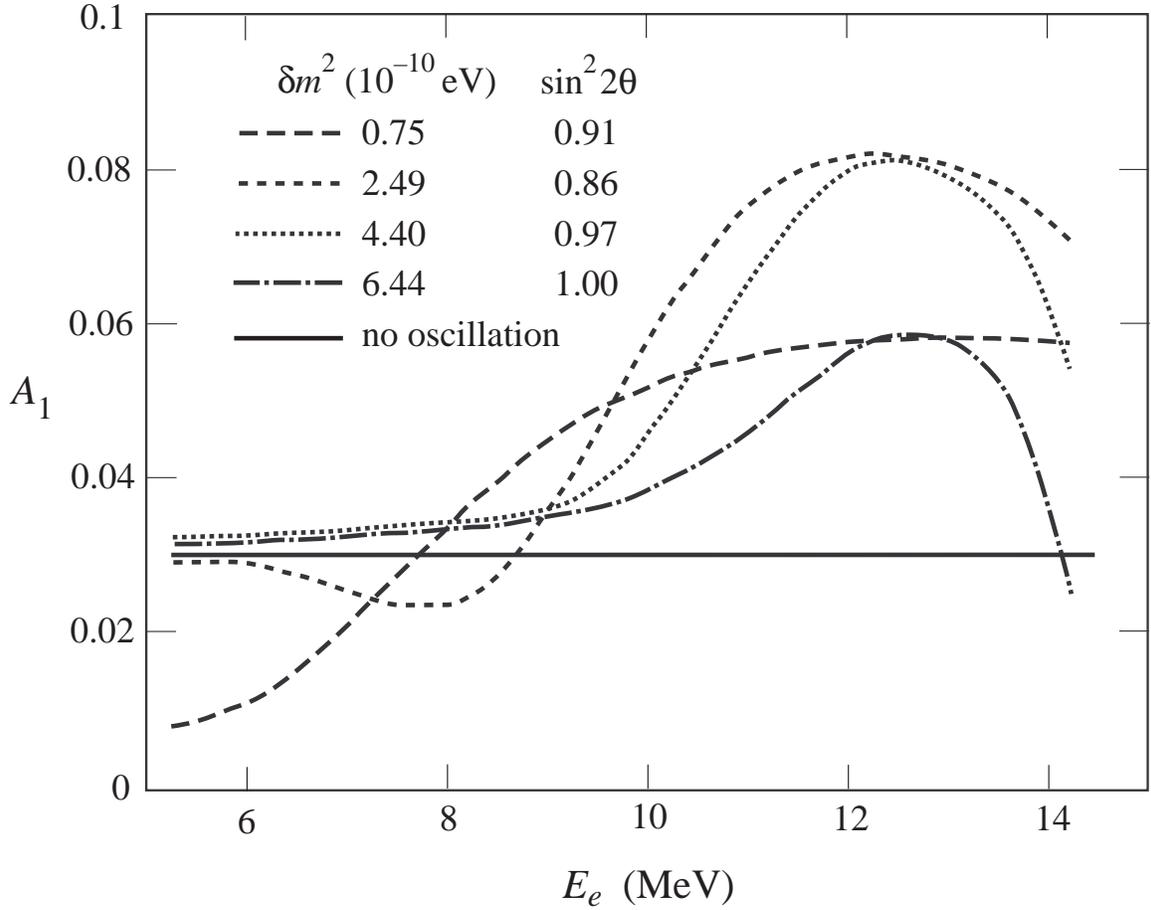}

\caption[]{\label{SuperKseas} Predicted values for the seasonal
asymmetry $A_1$ defined in Eq.~(\ref{asy1}) versus electron energy in
the SuperK experiment for the four vacuum neutrino oscillation solutions
A (long-dashed curve), B (short-dashed), C (dotted), and D
(dash-dotted) listed in Table~\ref{sunfits}. Also shown is the
energy-independent value for no oscillations (solid).}

\end{figure}

In order to extract the most information from the SuperK data it is
advantageous to plot the seasonal asymmetries versus the observed
electron energy. In Fig.~\ref{SuperKseas} we show $A_1$ versus $E_e$ for
Solutions~A, B, C, and D of Table~\ref{sunfits}. Although the
asymmetries are not large, each solution has a characteristic shape,
especially Solution~A. The energy dependence of $A_1$ clearly
distinguishes the oscillation scenarios from the asymmetry induced only
by the seasonal flux variation.  Similar measurements can be done in the
SNO experiment~\cite{SNO}. The deviation of $A_2$ from the value for no
oscillations is at most about 0.003, and does not provide a good
discrimination between models.

\begin{figure}
\centering\leavevmode
\epsfysize=5.5in\epsffile{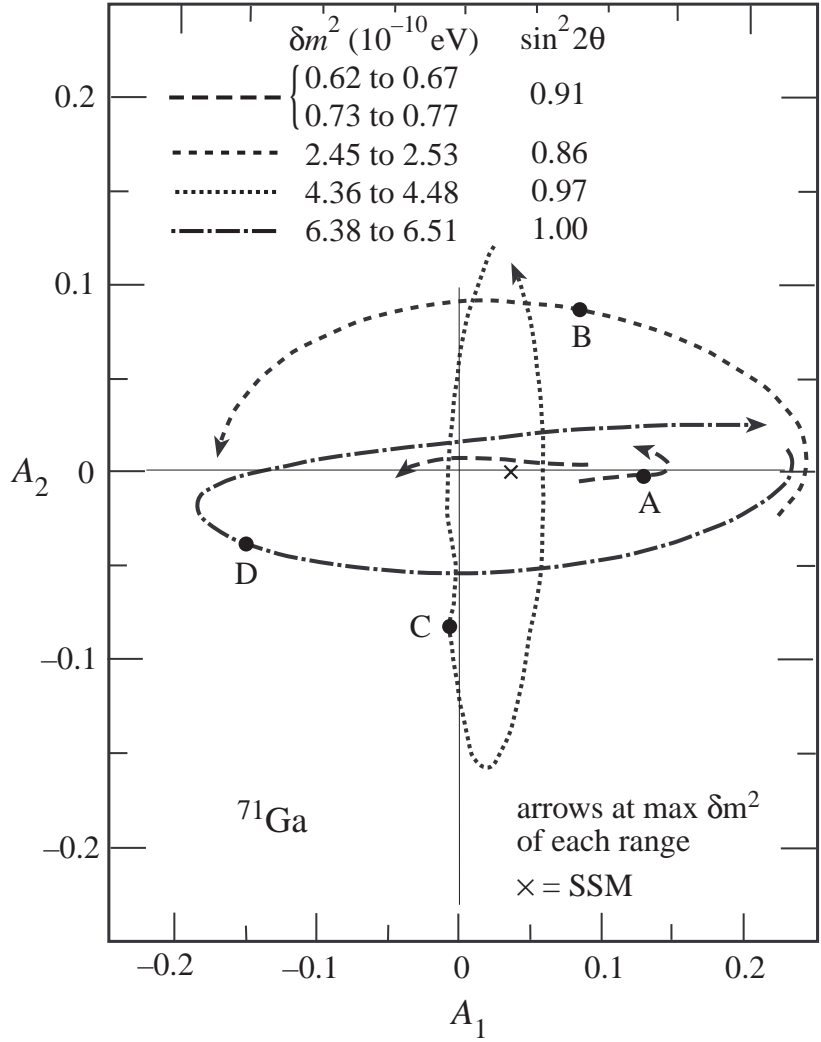}

\caption[]{\label{a1a2ga} Typical predicted seasonal asymmetries $A_2$
versus $A_1$ (defined in Eqs.~(\ref{asy1}) and (\ref{asy2})) in the
GALLEX and SAGE experiments for solutions from the four ``finger''
regions of Fig.~\ref{2nusun}: (a) $\sin^22\theta_3 = 0.74$, $\delta
m^2_{sun} = (0.62-0.67)\times10^{-10}$~eV$^2$ and $\sin^22\theta_3 =
0.91$, $\delta m^2_{sun} = (0.73-0.77)\times10^{-10}$~eV$^2$
(long-dashed curves), (b) $\sin^22\theta_3 = 0.86$, $\delta m^2_{sun} =
(2.45-2.53)\times10^{-10}$~eV$^2$ (short-dashed), (c) $\sin^22\theta_3
= 0.97$, $\delta m^2_{sun} = (4.36-4.49)\times10^{-10}$~eV$^2$ (dotted),
and (d) $\sin^22\theta_3 = 1.00$, $\delta m^2_{sun} =
(6.38-6.51)\times10^{-10}$~eV$^2$ (dash-dotted). The predictions of
best-fit solutions A, B, C, D from Table~\ref{sunfits} are indicated by
the solid circles. In cases (c) and (d), $\delta m^2_{sun}$ can vary
over a wider range than shown; the additional values give orbits in
$A_1$--$A_2$ space that are slightly shifted from those shown here. The
arrow on each curve is at the maximum value of $\delta m^2_{sun}$ for
each range. The no oscillation prediction (where the only variation in
signal comes from the variation of the flux due to the elliptical orbit)
is shown by a cross.}

\end{figure}

The GALLEX and SAGE $^{71}$Ga experiments may also exhibit a seasonal
variation with increased statistics if there are vacuum oscillations of
solar neutrinos~\cite{bpw,rosen}. In Fig.~\ref{a1a2ga} we plot the
predictions for $A_2$ versus $A_1$ in the GALLEX and SAGE experiments
for a range of $\delta m^2_{sun}$ from each of the four ``finger''
regions in Fig.~\ref{2nusun}. Most of the seasonal asymmetry in the
$^{71}$Ga experiments is due to the monoenergetic $^7$Be neutrinos which
constitute about 25\% of the signal in the SSM.

The BOREXINO experiment~\cite{BOREXINO} will primarily measure the
$^7$Be neutrinos ($E_\nu = 0.862$ MeV) using the process
$\nu e \rightarrow \nu_e$. The final-state electron kinetic energy
$T_e$ has a maximum value of 0.665~MeV for $^7$Be neutrinos. The $pp$
neutrinos have a maximum $T_e$ of 0.26~MeV, and also there is a
considerable background for $T_e < 0.25$~MeV. Therefore selecting events
in the range $0.26 {\rm~MeV} \le T_e \le 0.665$~MeV will preferentially
select the $^7$Be neutrino signal. Neutrinos from the $pep$ reaction and
the CNO cycle will also give final-state electrons in this energy
range, but the $^7$Be neutrinos represent more than 80\% of the signal,
assuming the SSM.

\begin{figure}
\centering\leavevmode
\epsfysize=5.5in\epsffile{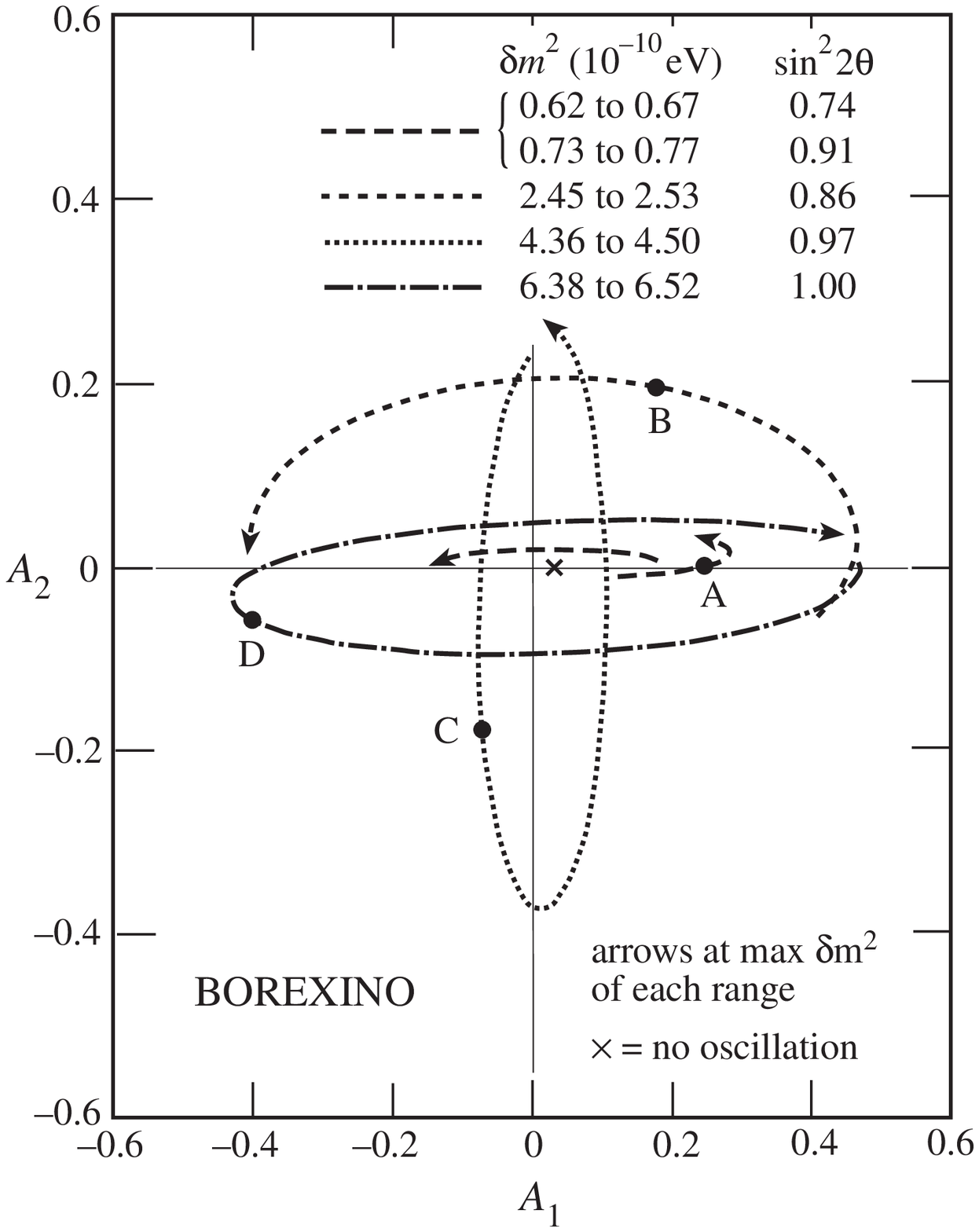}

\caption[]{\label{a1a2b} Same as Fig.~\ref{a1a2ga} for the BOREXINO
experiment. Cross sections have been calculated for final-state
electron kinetic energies in the range $0.26 {\rm~MeV} \le T_e \le
0.665$~MeV.}

\end{figure}

Predictions for event rate and seasonal variation in the BOREXINO
experiment for various vacuum oscillation parameters have previously
been discussed in the literature~\cite{seasonal1,bpw}. Here we examine
how the seasonal asymmetries in Eqs.~(\ref{asy1}) and (\ref{asy2}) may
be used to further discriminate the different vacuum neutrino solutions
in Table~\ref{sunfits}. In Fig.~\ref{a1a2b} we plot the predictions for
$A_2$ versus $A_1$ in the BOREXINO experiment for a range of $\delta
m^2_{sun}$ from each of the four ``finger'' regions in
Fig.~\ref{2nusun}. The seasonal asymmetries are potentially larger than
in the $^{71}$Ga case since the $^7$Be neutrinos are a much larger
fraction of the signal. Although there are some regions where the
predictions of two or more solutions overlap, combining the asymmetry
information with the event rate should significantly reduce the allowed
parameter regions for the vacuum solutions, and could select the
appropriate ``finger'' region.

\bigskip
\section{Acknowledgements}

VB thanks Fermilab for support as a Frontier Fellow and for kind
hospitality.  We thank John Learned for stimulating discussions
regarding the Super-Kamiokande atmospheric data, and we thank Sandip
Pakvasa and Tom Weiler for collaboration on previous related work. We
are grateful to Todor Stanev for providing his atmospheric neutrino flux
program.  This work was supported in part by the U.S. Department of
Energy, Division of High Energy Physics, under Grants
No.~DE-FG02-94ER40817 and No.~DE-FG02-95ER40896, and in part by the
University of Wisconsin Research Committee with funds granted by the
Wisconsin Alumni Research Foundation.

\newpage


\begin{table}
\caption[]{Best-fit two-neutrino solutions to the atmospheric data for
different oscillation scenarios. The $\sin^22\theta$ in each case
corresponds to the effective two-neutrino oscillation amplitude, with
$\theta=\theta_2$ ($\theta_1$) for $\nu_\mu \rightarrow \nu_\tau$
($\nu_\mu \rightarrow \nu_e$) oscillations.}

\label{atmfits}
\vspace{0.5 cm}
\vbox{\footnotesize
\tabcolsep=.5em
\centering\leavevmode
\begin{tabular}{|c|ccc|c|c|}
\hline
Oscillation & $\delta m^2_{atm}$ (10$^{-3}$~eV$^2$) & $\sin^22\theta$
& $\alpha$ & $\chi^2_{tot}$/DOF & Goodness-of-fit\\ 
\hline
$\nu_\mu \rightarrow \nu_\tau$ & 2.8 & 1.00 & 1.16 & 7.1/13 & 90\% \\
$\nu_\mu \rightarrow \nu_\tau$ & 1.1 & 0.85 & 1.00 (fixed) & 22.8/14 & 6\% \\
$\nu_\mu \rightarrow \nu_e$ & 1.1 & 0.88 & 0.71 & 81.9/13 & $4\times10^{-12}$\\
\hline
\end{tabular} }
\end{table}


\begin{table}
\caption[]{Best-fit two-neutrino solutions to the solar data in the four
``finger'' regions of $\delta m^2_{sun}$ depicted in Fig.~\ref{2nusun}.
The top half of the table corresponds to a fit to all solar data, while
the bottom half shows the results for a fit with the $^{37}$Cl data
excluded. The $\chi^2$ sums do not add to the total in some cases due to
rounding.}

\label{sunfits}
\vspace{0.5 cm}
\vbox{\footnotesize
\tabcolsep=.5em
\centering\leavevmode
\begin{tabular}{|c|ccc|ccc|c|c|}
\hline
Solution
&$\delta m^2_{sun}$ (10$^{-10}$~eV$^2$) & $\sin^22\theta_3$ & $\beta$
& $^{37}$Cl & $^{71}$Ga & Super-K & $\chi^2_{tot}$ & Goodness-of-fit \\ 
\hline
A & 0.75 & 0.91 & 1.62 & 0.8 & 1.5 & 19.4 & 21.6 & 16\% \\
B & 2.49 & 0.86 & 0.84 & 5.9 & 2.2 & 18.2 & 26.3 &  5\% \\
C & 4.40 & 0.97 & 0.80 & 6.5 & 4.7 & 12.3 & 23.5 & 10\% \\
D & 6.44 & 1.00 & 0.80 & 7.4 & 5.3 & 15.0 & 27.6 &  4\% \\
\hline
           & 0.75 & 0.85 & 1.38 & --- & 1.1 & 19.0 & 19.0 & 17\% \\
           & 2.47 & 0.77 & 0.78 & --- & 1.3 & 16.0 & 16.0 & 30\% \\
           & 4.35 & 0.97 & 0.80 & --- & 1.2 & 12.1 & 12.1 & 58\% \\
           & 6.35 & 0.97 & 0.78 & --- & 1.3 & 15.2 & 15.2 & 35\% \\
\hline
\end{tabular} }
\end{table}

\clearpage


\vfill
\eject

\clearpage


\begin{thebibliography}{99}
\bibitem{stanevflux}
We use an updated version of the neutrino
flux in the first paper of Ref.~\cite{flux};
a program to calculate the flux was provided by T. Stanev.
\end{thebibliography}
\end{document}